\documentclass[aip,reprint]{revtex4-1}
\usepackage[colorlinks=true,linkcolor=blue,citecolor=blue,urlcolor=blue]{hyperref}
\usepackage{graphicx}
\usepackage[utf8]{inputenc}
\usepackage[T1]{fontenc}
\usepackage{etoolbox}
\usepackage{lineno}
\usepackage[percent]{overpic}
\usepackage{float}
\usepackage{xcolor}
\usepackage{svg}
\usepackage{tikz}
\usepackage{amsmath}
\usepackage{orcidlink}
\usepackage{lineno}
\usetikzlibrary{decorations.pathmorphing}
\usepackage{xcolor}
\usepackage{soul} 
\newcommand{\rev}[1]{{\color{black}#1}}
\draft 
\newcommand{\UniRoma}{Dipartimento di Fisica - Sapienza Università di Roma - Piazzale Aldo Moro 2, 00185, Roma, Italy}
\newcommand{\INFNRoma}{INFN - Sezione di Roma - Piazzale Aldo Moro 2, 00185, Roma, Italy}
\newcommand{\UniGre}{Univ. Grenoble Alpes, CNRS, Grenoble INP, Institut Néel, 38000 Grenoble, France}
\begin{document}

\title{Enhanced Athermal Phonon Responsivity in a Kinetic Inductance Detector with Integrated Phonon Collectors} 


\author{L.~Pesce\,\orcidlink{0009-0001-5659-4691}}\email{leonardo.pesce@uniroma1.it}
\affiliation{\UniRoma}\affiliation{\INFNRoma}
\author{A.L.~De~Santis\,\orcidlink{0009-0005-4288-3758}}\altaffiliation{now at Gran Sasso Science Institute (GSSI), 67100, L’Aquila, Italy}\affiliation{\UniRoma}\affiliation{\INFNRoma}
\author{M.~Calvo\,\orcidlink{0000-0002-8752-6325}}\affiliation{\UniGre}
\author{M.~Cappelli\,\orcidlink{0009-0002-6148-5964}}\affiliation{\UniRoma}\affiliation{\INFNRoma}
\author{U.~Chowdhury}\altaffiliation{now at IQM Finland Oy, Espoo, Finland}\affiliation{\UniGre}
\author{A.~Cruciani\,\orcidlink{0000-0003-2247-8067}}\affiliation{\INFNRoma}
\author{G.~Del~Castello\,\orcidlink{0000-0001-7182-358X}}\affiliation{\INFNRoma}
\author{D.~Delicato\,\orcidlink{0009-0005-0516-6872}}\affiliation{\UniGre}\affiliation{\UniRoma}\affiliation{\INFNRoma}
\author{M.~Folcarelli\,\orcidlink{0009-0009-7799-2515}}\affiliation{\UniRoma}\affiliation{\INFNRoma}
\author{M.~del~Gallo~Roccagiovine\,\orcidlink{0009-0006-5861-7443}}\affiliation{\UniRoma}\affiliation{\INFNRoma}
\author{A.~Monfardini\,\orcidlink{0000-0001-5337-5533}}\affiliation{\UniGre}
\author{D.~Quaranta\,\orcidlink{0009-0000-2954-4456}}\affiliation{\UniRoma}\affiliation{\INFNRoma}
\author{M.~Vignati\,\orcidlink{0000-0002-8945-1128}}\affiliation{\UniRoma}\affiliation{\INFNRoma}
\begin{abstract}
Cryogenic phonon detectors are adopted in light dark matter searches and coherent elastic neutrino-nucleus scattering experiments as they can achieve low energy thresholds. The phonon mediated sensing of silicon particle absorbers has already been proved with Kinetic Inductance Detectors (KIDs), acting both as sensors and athermal phonon absorbers. In this work we present the design and the performance of an improved \rev{detector design}\rev{. In this architecture, the} KID acts only as sensor and is coupled to dedicated phonon collectors. \rev{When a signal is coming from the substrate, the presence of a separated collector allows to detect an higher increase of quasi-particles density}, thereby enhancing its responsivity. The meander of the KID is composed of a $77 \,\mathrm{nm}$ trilayer wire of Aluminum-Titanium-Aluminum, while the phonon collectors are made of a $100\,\mathrm{nm}$ Aluminum layer and act as quasi-particles funnels. Inside the collectors, the absorbed athermal phonons generate quasi-particles which, after diffusion, are trapped in the lower-gap superconducting trilayer. The performance of this setup is compared to that of a standard phonon-mediated KID, showing an increase\rev{d phonon collection efficiency by a factor of around 7}.
\end{abstract}

\maketitle

\begin{center}
\footnotesize
Copyright 2026 Leonardo Pesce et al. This article is distributed under a Creative Commons Attribution (CC BY) License. This article appeared in Applied Physics Letters and may be found at \url{https://doi.org/10.1063/5.0323811}
\end{center}

Experiments searching for direct light dark matter interaction~\cite{cresst, edelwiss_III, Soudan} or coherent elastic neutrino-nucleus scattering~\cite{gram_scale, Chooz, ricochet, miner, coherent, conus} address two main challenges: the need for high sensitivity to nuclear recoils and the scaling of the target mass. 

\rev{Small} nuclear recoil energies require detectors with energy thresholds as low as possible. For this reason, cryogenic detectors~\cite{cresst, edelwiss_III, gram_scale, ricochet, cdms_resolution} are adopted as phonon sensors coupled to target crystals, \rev{achieving} energy thresholds of the order of 1-10~eV~\cite{sub_ev, gram_scale, cdms_resolution}. At the same time, the development of mass-scalable instruments is essential for achieving improved sensitivity to increasingly smaller dark-matter interaction cross sections, assuming a Weakly Interacting Massive Particle (WIMP) model~\cite{dm_review}. \rev{Low threshold } experiments with target mass of tens of grams have been built~\cite{cresst, gram_scale, edelweiss_gram, cdms_resolution}, but scaling-up to the kilogram still remains an experimental challenge.

The BULLKID project developed a particle detector designed for tackling the mass scaling problem~\cite{bullkid_generico}. It consists of a monolithic array of 60 silicon dice, each one with mass of 0.34\,g and volume $5.4\times5.4\times5\,\mathrm{mm^3}$, carved from a wafer with a diameter of 7.6\,cm and 5\,mm thickness. The dice are sensed with 60 KIDs acting as phonon-mediated particle detectors. The best energy resolution achieved amounts to 27\,eV \rev{corresponding to} an \rev{energy} threshold of 160\,eV~\cite{bullkid_fondo}, still far from the current state of art of  other cryogenic detectors. For this reason, it is important to develop new sensors \rev{with lower} energy threshold. In this letter, we investigate a KID design aimed at achieving this goal.

KIDs,  originally developed for the detection of optical~\cite{kid_optical} and millimeter-wavelength~\cite{astro_kid_mm} light, have found application also as phonon-mediated particle detectors~\cite{bullkid_generico, phonon_imaging,kid_x,kid_dm}. A KID is a superconducting LC resonator~\cite{kid_design} in which the total inductance $L$ is the sum of the magnetic inductance $L_g$ and the kinetic inductance $L_k$. When \rev{an} energy $E$ is released in the \rev{KID's substrate}, athermal phonons are generated and absorbed by the KID, where they can break Cooper Pairs (CPs) releasing Quasi-Particles (QPs). This process reduces the CP number density, which in turn increases the kinetic inductance $L_k$, leading to a resonant frequency shift, $\Delta f_r$, and to a reduction of the resonance depth~\cite{kid_design,mazin2005mkid,modello1,modello2,non_linearity}.

The change of the CP density is detected by monitoring the phase and magnitude modulation of the complex single pole transmission function $S_{21}(y)$ of the resonator~\cite{KID_germanium}:
\begin{equation}\label{eq:S21}
    S_{21}(y) = 1-\frac{Q}{Q_c} \frac{1}{1+2jy}
\end{equation}
 where $Q = (Q_c^{-1} + Q_i^{-1})^{-1}$ is the total quality factor of the resonator, which combines the coupling quality factor $Q_c$ and the internal quality factor $Q_i$~\cite{kid_design}. Finally, the parameter $y = Q\cdot\Delta f_r/f_r$ is the detuning of the readout frequency relative to the resonance width~\cite{kid_design, mazin2005mkid, non_linearity}. Usually, the phase readout $\varphi$ provides a better signal-to-noise ratio~\cite{kid_design}; therefore \rev{it} will be adopted in this work.

The signal phase responsivity $r$ of a KID is~\cite{kid_design, mazin2005mkid, resp, temperature_scan, KID_germanium}:
\begin{equation}\label{eq_responsivity}
        r =  \frac{\eta}{V_{\mathrm{KID}}} \frac{\alpha S_2(f_r, T)Q}{N_0\Delta_0^2} \frac{1}{1+4y^2}
\end{equation}
where $\eta$ is the energy to pair-breaking conversion efficiency, $V_{\mathrm{KID}}$ is the active inductor volume, $\alpha = L_k/(L_k+L_g)$ is the kinetic inductance fraction, and $S_2(f_r, T)$ is a dimensionless factor of order one from Mattis-Bardeen theory. Finally, \rev{$N_0$ is the single-spin volume density of electron states at the Fermi energy} and $\Delta_0 \simeq 1.76k_BT_c$ is the superconducting gap, with $k_B$ the Boltzmann constant and  $T_c$ the critical temperature. We emphasize that in Eq.~\ref{eq_responsivity} the term $\Delta_0$ appears quadratically as the product of two identical gap factors: one term arises from the number of CPs broken by a given energy deposition $E$, scaling as $E/\Delta_0$, while the second originates from the single–spin volume density, $N_0\Delta_0$, within the sensor~\cite{kid_design, mazin2005mkid, resp, temperature_scan}.

If phonons absorption is not saturated and losses due to supports or interfaces do not dominate, the efficiency $\eta$ scales almost linearly with the \rev{surface in contact with the substrate $A_\mathrm{ph}$~\cite{eta_phonon, phon_rec, desantis2022phonon}:
\begin{equation}\label{eq:eta}
    \eta \propto A_{\mathrm{ph}}
\end{equation}

In a standard phonon-mediated KID, the collection phonon surface $A_{\mathrm{ph}}$ coincides with the active contact surface of the inductor $A_{\mathrm{KID}}$. Increasing such surface coverage does not lead to an increase of the KID signal due to the consequent increase of the inductor volume $V_{\mathrm{KID}}$, as follows from Eq.~\ref{eq_responsivity}. Conversely, separating these surfaces allows to design devices with large $A_\mathrm{ph}$ and small $A_\mathrm{KID}$ (thus, small $V_\mathrm{KID}$), which would enhance the detector responsivity.}

\begin{figure}[t]
    \centering
    \includegraphics[width=\columnwidth]{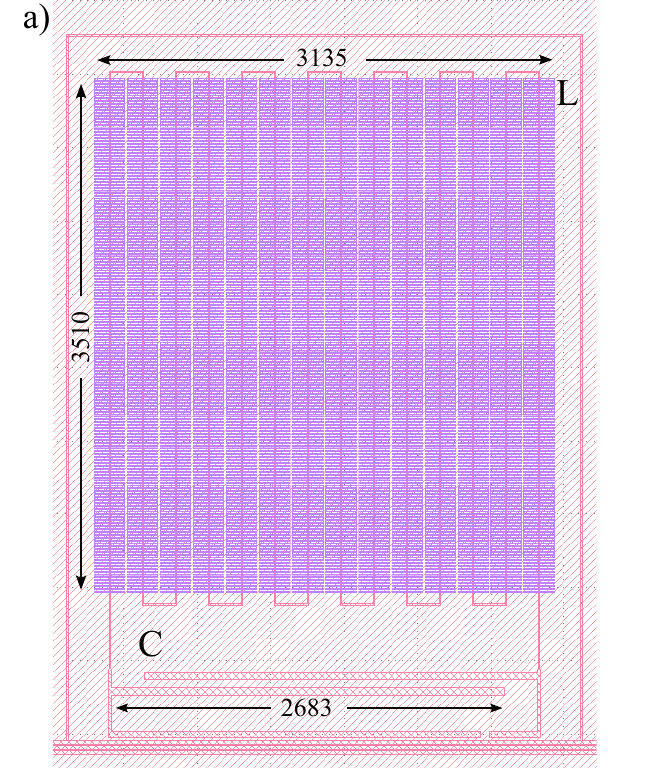}  
    \includegraphics[width=\columnwidth]{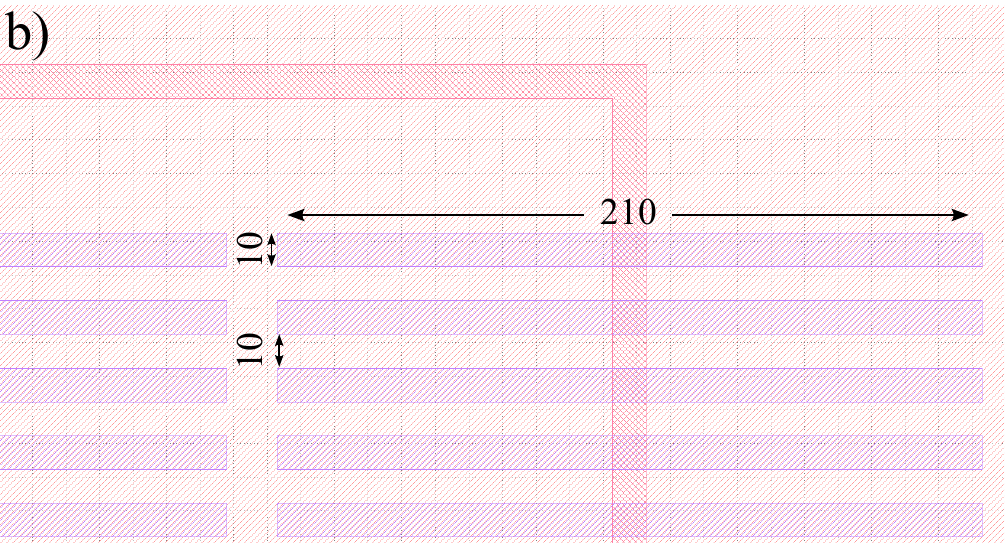}
    \caption{(a) Layout of the FunKID. The funnels in purple intersect the inductive meander indicated with L, while the capacitor is indicated with C. (b) Detail of the funnels. The measures are in $\mu\mathrm{m}$.}
    \label{fig:kid_funnel+funnel}
\end{figure}

\rev{Following the same concept of previous works made with TESs~\cite{tes_1, tes_2, cdms_resolution},} we developed and tested a phonon-mediated KID with separate phonon collectors~\cite{desantis2022phonon, mazin2006assorbitore, procceding, qp_trapping_model} that \rev{also act} as QPs funnels (see Fig.~\ref{fig:kid_funnel+funnel}). We refer to this device as the FunKID and to the phonon collectors as the funnels. The resonator is \rev{patterned out} of a 77\,nm trilayer of Aluminum-Titanium-Aluminum (AlTiAl), with 14\,nm-33\,nm-30\,nm thicknesses, respectively\rev{, which already exhibited a $Q_i> 0.5\,\mathrm{M}$~\cite{altial}}. The inductive part of the KID is composed of seven \rev{segments} of around $3600\,\mathrm{\mu m}\times10\,\mathrm{\mu m}$, evenly spaced by $215 \,\mathrm{\mu m}$ (Fig.~\ref{fig:kid_funnel+funnel}a). The funnels are Aluminum structures with $210\,\mathrm{\mu m}\times10\,\mathrm{\mu m} \times 100\,\mathrm{nm}$ volume intersecting the inductor, evenly spaced by $10\, \mathrm{\mu m}$ (Fig.~\ref{fig:kid_funnel+funnel}b). \rev{The total phonon collection surface, which coincides with the total contact surface of the funnels with the substrate, is $A_{\mathrm{ph}}\simeq5.0\,\mathrm{mm}^2$. The surface of the resonator in contact with the substrate is $A_{\mathrm{KID}}\simeq 0.55 \,\mathrm{mm}^2$, giving a surface ratio of $A_{\mathrm{ph}}/A_{\mathrm{KID}}=9$.}

Due to proximity effects~\cite{altial,bilayer,zhao2017,multilayer_general} among the two Aluminum layers ($T_c\simeq 1.2\,\mathrm{K}$~\cite{zhao2017}) and the central Titanium layer ($T_c\simeq0.4\,\mathrm{K}$~\cite{zhao2017}), the AlTiAl sensor has an intermediate critical temperature $T_c$, which is lower than that of the funnels, made of Aluminum. For the same design thicknesses, the CALDER project measured the critical temperature of an AlTiAl trilayer as $T_c \simeq 0.81\,\mathrm{K}$~\cite{altial, calder}\rev{. A} similar value is expected for this resonator, assuming comparable \rev{interface} effects~\cite{zhao2018calculation}.

\begin{figure}[t]
    \centering
    \includegraphics[width=\columnwidth]{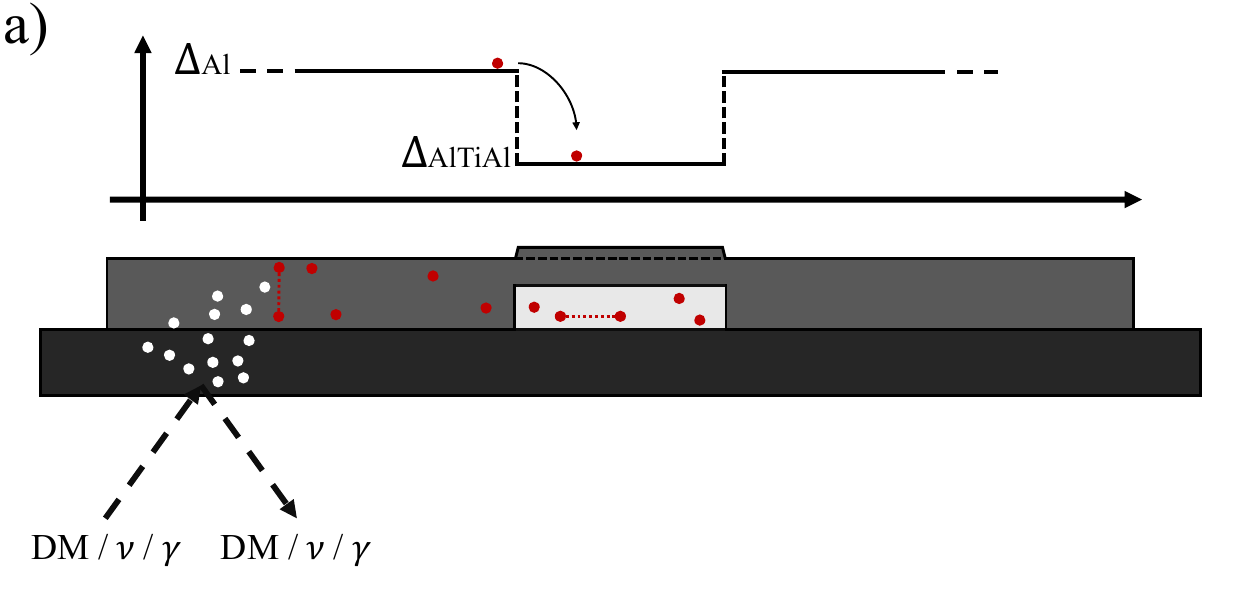}  
    \includegraphics[width=\columnwidth]{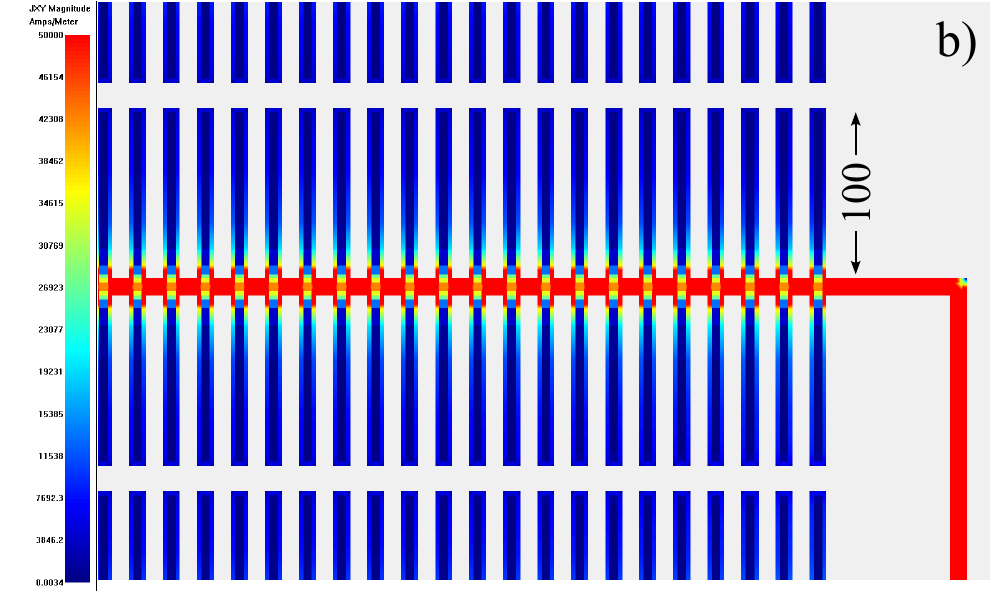}
    \caption{(a) Detection scheme of the FunKID. The white dots represent the athermal phonons generated by an energy deposition in the substrate, the red paired dots are the CPs and the single red dots the QPs diffusing from the funnel to the KID. More details in the text. (b) Current density simulated with SONNET. The length reported is measured in $\mathrm{\mu m}$.}
    \label{fig:funnel_sketch+current}
\end{figure}

In the FunKID design, schematically reported Fig.~\ref{fig:funnel_sketch+current}a, the phonons produced in the substrate are mostly absorbed by the funnels, where they can break CP generating QPs which, through diffusion, reach the sensor. Due to the energy gap difference between funnels with gap $\Delta_{\mathrm{Al}}$ and sensor with gap $\Delta_{\mathrm{AlTiAl}}$, QPs with energy $\Delta_{\mathrm{Al}}$ relax down to $\Delta_{\mathrm{AlTiAl}}$ and are trapped within the sensor, preventing their recombination in the funnels~\cite{tes_1, tes_2, mazin2006assorbitore, qp_trapping_model, qp_trapping}. \rev{We emphasize that the trapping efficiency depends on the gap difference between the funnels and collectors. The decision to operate with such a small difference is based on the need to maintain sufficiently high $Q_i$, even if it may lead to lower trapping efficiency. However, determining the exact trapping efficiency is non-trivial and would require dedicated theoretical models~\cite{tes_1, qp_trapping, trapping_model2, al_ox}.}

The length of the funnels \rev{was chosen to be} smaller than the characteristic QP diffusion length in Aluminum, 
$\lambda_r = \sqrt{D\tau_r}$~\cite{diff_model}, where $D$ is the QP diffusion coefficient and $\tau_r$ the QP recombination time~\cite{rec_time_Al, rec_time_Al2}. 
We adopted $D = 22.5\,\mathrm{cm^2/s}$~\cite{D_qp}, \rev{while on previous resonators in $60\,\mathrm{nm}$ Al ($\mathrm{RRR}\simeq 4$) we measured $\tau_r\simeq 3\,\mathrm{ms}$~\cite{bullkid_generico} at base temperature, yielding a diffusion length of 
$\lambda_r \simeq 2.6\,\mathrm{mm}$, much higher than the funnels length of $200\,\mathrm{\mu m}$}. We can also define a characteristic time scale quantifying the maximum diffusion time of QPs inside the funnels before reaching the sensor, 
$\tau_{\mathrm{diff}} \sim l^2/D = 4\,\mathrm{\mu s}$, 
where $l = 100\,\mathrm{\mu m}$ is half of the funnel length. This sets the minimum timescale over which the sensor needs to integrate the signal.

\rev{The design of the funnels reasonably limits the current density leaking into them}, as shown in the simulation performed with the SONNET software~\cite{sonnet2025} reported in Fig~\ref{fig:funnel_sketch+current}b. In particular, \rev{the current in the sensor at resonance} flows through the inductive meander, while only a small fraction leaks in the funnels. However, we notice that possible limitations of the simulations may arise from the \rev{lack of} modeling of the overlap region between the funnels and the resonator.

We also notice that in the FunKID the separation between phonon absorbers and sensor modifies the term $\Delta_0^2$ of Eq.~\ref{eq_responsivity}. In this design, the CPs breaking occurs in the funnels, with gap $\Delta_{\mathrm{Al}}$, while the QPs \rev{recombination} refers to that \rev{in} the sensor, with gap $\Delta_{\mathrm{AlTiAl}}$. Therefore, the term $\Delta_0^2$ is replaced by $\Delta_{\mathrm{Al}} \Delta_{\mathrm{AlTiAl}}$.

\rev{The FunKID was designed with a coupling quality factor $Q_c=20\,\mathrm{k}$ and resonant frequency $f_r = 1009.6\,\mathrm{MHz}$. The small $Q_c$ design was chosen to ensure a deep resonance even in the case of degraded internal quality factor $Q_i$. Conversely, it is worth to highlight that choice limits the ring time to $\tau_{\mathrm{ring}} = Q/\pi f_r = 5\,\mathrm{\mu s}$~\cite{bullkid_generico}. Since $\tau_{\mathrm{ring}}$ corresponds to the integration time of the detector, a shorter $\tau_\mathrm{ring}$ than $\tau_\mathrm{diff}$ and phonon arrival time $\tau_\mathrm{ph}$~\cite{eta_phonon} could result in a degradation of the FunKID response.} 

\rev{Starting from Eq.~\ref{eq_responsivity}, we can write the ratio of the efficiency of the FunKID $\eta_\mathrm{F}$ due to the presence of the funnels, relative to an identical KID without the funnels with efficiency $\eta_\mathrm{K}$ and phonon collection surface $A_\mathrm{KID}$:
\begin{equation}\label{eq:gain}
    \frac{\eta_\mathrm{F}}{\eta_\mathrm{K}}\simeq \left(\frac{r_{\mathrm{FunKID}}}{r_{\mathrm{KID}}}     \frac{\alpha_{\mathrm{KID}}}{\alpha_{\mathrm{FunKID}}}
    \frac{Q_{\mathrm{KID}}}{Q_{\mathrm{FunKID}}}-1\right) 
    \frac{\Delta_{\mathrm{Al}}}{\Delta_{\mathrm{AlTiAl}}}
\end{equation}
where we assume that $N_0$ and $S_2(f_r, T)$ are approximately equal between the two resonators and a linear response,  with $1+4y^2\simeq 1$. The second therm in parentheses arises from the contribution of the FunKID's resonator to the total device responsivity, acting as a standard KID with gap $\Delta_\mathrm{AlTiAl}$. In order to evaluate the real effect of the funnels on the collection efficiency, we should therefore subtract this contribution from the total responsivity of the FunKID.}
 
\begin{figure}[t]
    \centering
    \includegraphics[width =\columnwidth]{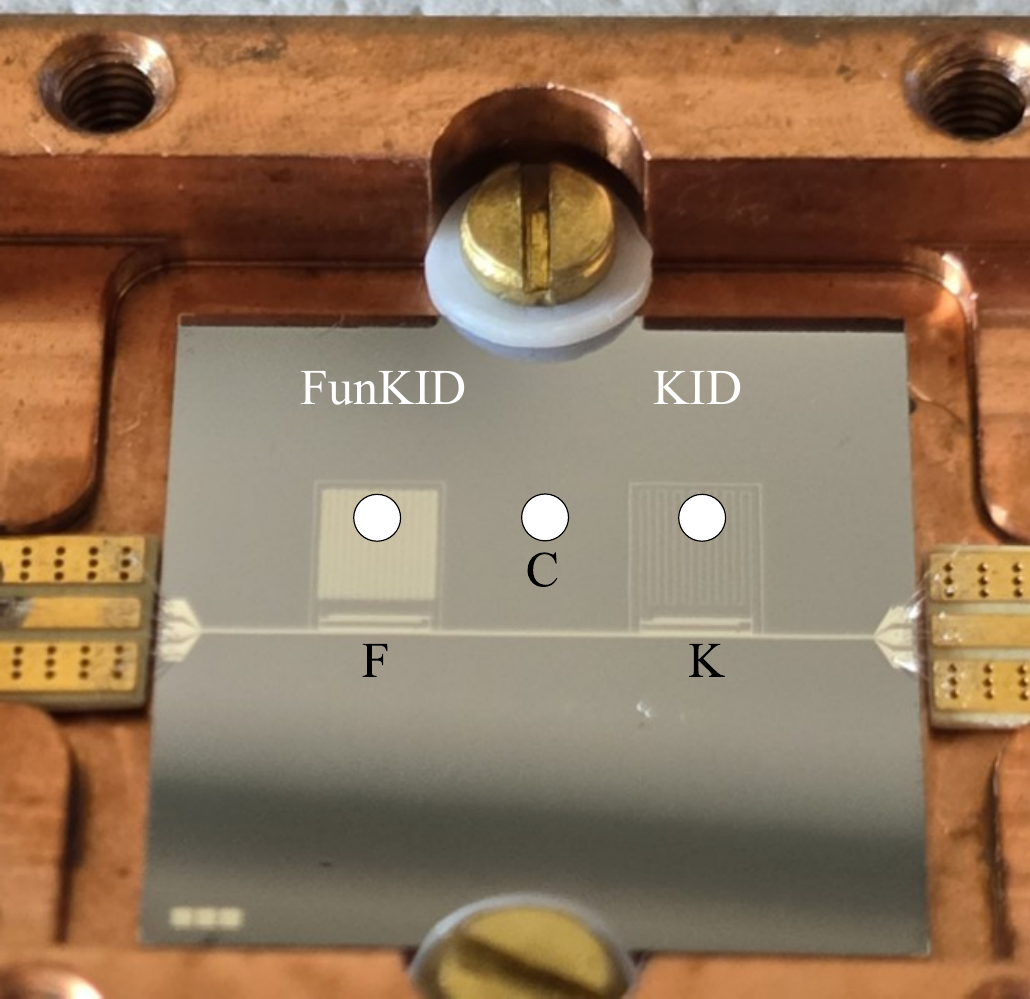}
    \caption{The two pixels on the silicon tile suspended by Teflon supports. On the left is the FunKID, while on the right the standard phonon-mediated KID. \rev{The white dots indicate the position of the fibers, and the letters correspond to their label.}}
    \label{fig:photo}
\end{figure}

The test device was fabricated at the PTA facility clean room in Grenoble~\cite{pta2025} using a $500\,\mathrm{\mu m}$ thick silicon tile. On the same substrate, an identical standard phonon-mediated KID, with a simulated resonant frequency $20\,\mathrm{MHz}$ higher than the FunKID, was also produced (Fig.~\ref{fig:photo}). The KID sensors and the feed-line were etched from the 77\,nm AlTiAl trilayer following a standard lithographic technique. After their fabrication, photo-resist was used to coat the surface of the wafer, and gaps in the shape of the funnels were patterned on it. \rev{After applying Ar ion etching to remove Al-oxide, the} funnels were then realized by selectively depositing $100\,\mathrm{nm}$ of Aluminum in the patterned gaps, following a lift-off process.

The tile was \rev{then} mounted into a copper holder by means of Teflon supports and wire bonded to $50\,\mathrm{\Omega}$ launchers. The device was operated in a dilution $^3\mathrm{He}/^4\mathrm{He}$ cryostat, at a base temperature $T_0 \simeq 35\,\mathrm{mK}$. Finally, a mu-metal shield was placed outside the cryostat to reduce magnetic-field effects~\cite{mu_metal}.

To calibrate the detectors, three optical fibers were placed facing the substrate backside: two were aligned with the resonators for their identification~\cite{procceding} and one was located at the center to calibrate \rev{both KIDs}. The fibers were pulsed with $400\,\mathrm{nm}$ photons bursts generated at room temperature by a controlled LED~\cite{lantern}.

The readout tones were generated by the NIXA superheterodyne radio-frequency detection system~\cite{NIXA_hardware, NIXa_firmware} transmitting and receiving tones at the KIDs resonant frequencies. The transmitted waves undergo an attenuation chain of  \rev{$56\,\mathrm{dB}$} before the chip, are transmitted past the resonators, then are amplified by 30\,dB by a Low Noise Amplifier~\cite{LNA} operated at 4\,K and finally received back by NIXA.

\begin{figure}[t]
    \centering
    \includegraphics[width=\columnwidth]{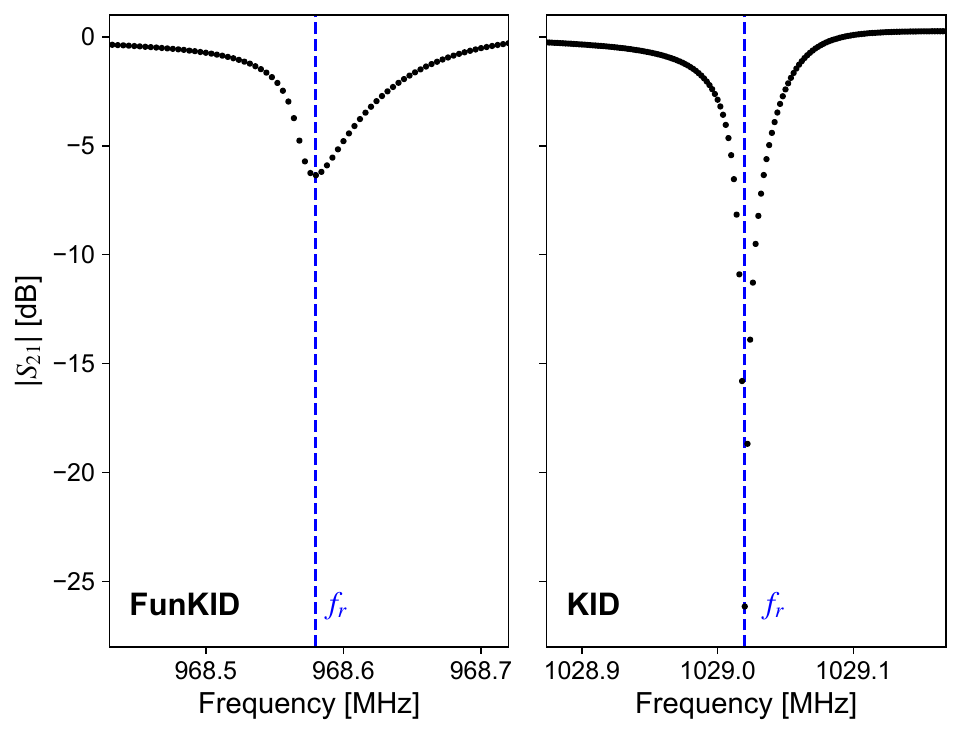}
    \caption{Transmission function amplitude $|S_{21}|$ around the resonance frequencies $f_r$ of the resonators.}
    \label{fig:fit}
\end{figure}

The resonances of the two KIDs are shown in Fig.~\ref{fig:fit}. We fitted both the amplitude of the transmission function $|S_{21}|$ and its real and imaginary components in order to extract the resonant frequencies and the quality factors $Q_c$, $Q_i$, and $Q$ of the two KIDs.  We also evaluated \rev{the} resonator ring times $\tau_{\mathrm{ring}} = Q/\pi f_r$. The results are reported in Tab.~\ref{tab:table1}.

\begin{table}[t]
  \centering
  \begin{tabular}{lcccccc}
    \hline
    \hline
    &$f_r\,\mathrm{[MHz]}$& $Q_c\,\mathrm{[k]}$ & $Q_i\,\mathrm{[k]}$ & $Q\,\mathrm{[k]}$ & $\tau_{\mathrm{ring}}\,\mathrm{[\mu s]}$\\
    \hline
    \bf{FunKID} & 968.6 & 19 & 20 & 10 & 3\\
    \bf{KID} & 1029.0 & 20 & 410 & 19 & 6\\
    \hline
    \hline
  \end{tabular}
  
  \caption{\label{tab:table1}Results of the resonant frequency $f_r$, quality factors $Q$, $Q_i$ and $Q_c$ and ring times $\tau_{\mathrm{ring}}$ of the two resonators.}
\end{table}

A shift of around $-40\,\mathrm{MHz}$ of the measured FunKID resonant frequency with respect to the simulation was observed, indicating that the overlap regions between the funnels and the resonator might have not been correctly modeled in the simulations. Conversely, the obtained KID resonant frequency is in agreement with simulation, assuming a kinetic inductance $L_k = 1.4\,\mathrm{pH/sq}$, which is the same of Ref.~\cite{altial}.  The coupling quality factors $Q_c\sim20\,\mathrm{k}$ are in line with the values simulated with SONNET, for both resonators. The internal quality factor of the FunKID is comparable to the coupling quality factor, \rev{currently limiting this design}. Conversely, the internal quality of the KID is higher than the FunKID and in line with the values obtained in Ref.~\cite{altial}, indicating that the reduced $Q_i$ of the FunKID is set by the funnels and not by fabrication processes of the trilayer AlTiAl film. \rev{While the demonstration of the FunKID's operation is not compromised by this limitation, the relevant contributions for the reduced $Q_i$ in the FunKID will be investigated in further works.}

For each resonator, we evaluated an estimate of kinetic inductance fraction $\alpha$, by combining the measured resonant frequencies $f_r$ with the resonant frequencies $f_0$ obtained from SONNET simulations performed with $L_k = 0$~\cite{altial}:
\begin{equation}
    \alpha = 1-\left(\frac{f_r}{f_0}\right)^2
\end{equation}

\rev{We obtained $\alpha = 18.9\,\%$ and $\alpha = 15.6\,\%$ for the FunKID and the KID, respectively. 
We compared these values with those obtained using the same approach with two devices identical to the tested samples, with the wire patterned out of $60\,\mathrm{nm}$ Al. The estimates for the Al devices yield $7.0\%$ and $5.0\%$ for the FunKID and KID, respectively, 
in agreement with an independent measurement of $\alpha$ obtained by fitting the resonator frequency shift as a function of temperature with the BCS model~\cite{temperature_scan} (for the details see the supplementary material).}

\rev{For the AlTiAl sample we} measured a feed-line $T_c = (0.83 \pm 0.08)\,\mathrm{K}$, corresponding to $\Delta_0 =  (125.9 \pm 12.6)\,\mathrm{\mu eV}$, in reasonable agreement with the $0.81\,\mathrm{K}$ reported in~\cite{altial}. We assumed that this value applies to both the resonators, although funnels likely modify the FunKID critical temperature, which cannot be measured directly in the presented setup. 

\begin{figure}[t]
    \centering
    \includegraphics[width=\columnwidth]{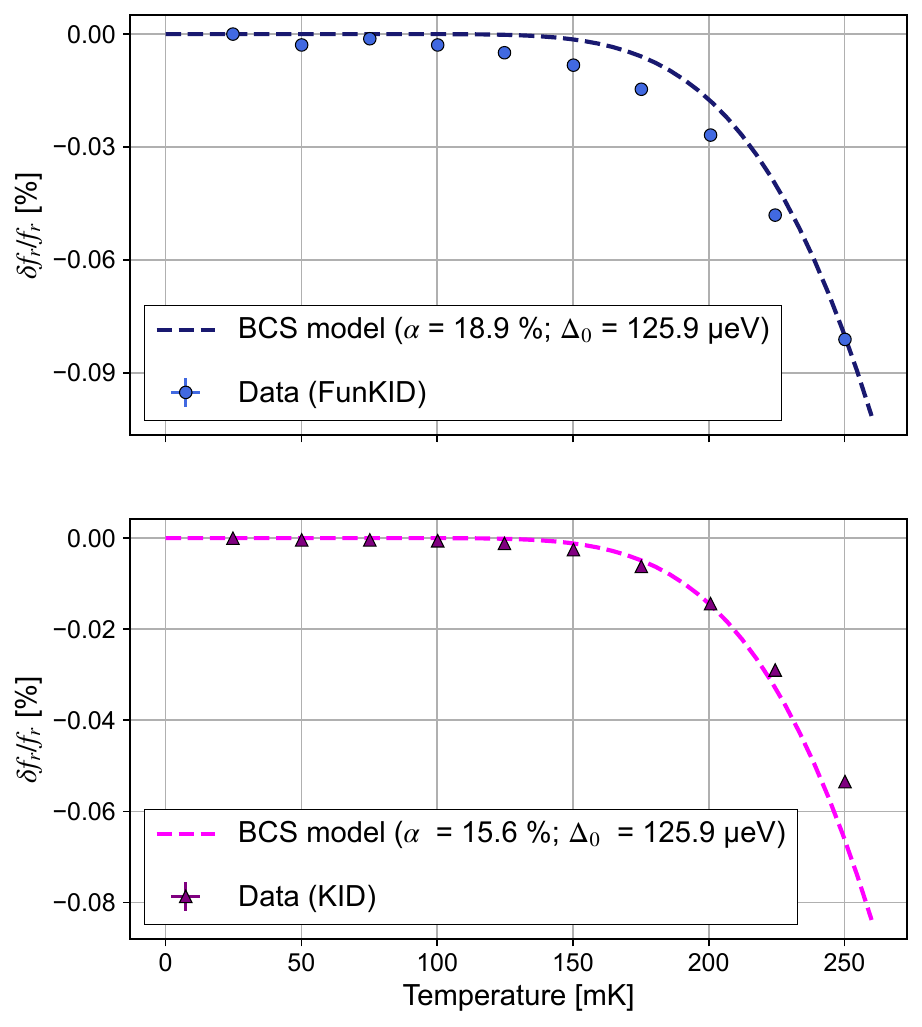}
    \caption{Relative shift of the resonant frequency with respect to the base temperature resonant frequency, $\delta f_r/ f_r$, for the FunKID in blue and KID in purple. The dashed curves are obtained with the BCS model calculated with our best \rev{estimates} of $\alpha$ and $\Delta_0$.}
    \label{fig:scan_T}
\end{figure}

\rev{To compare the data to a model based on the BCS theory~\cite{temperature_scan}, we measured the resonant frequency shift relative to the base temperature resonant frequency, $\delta f_r/ f_r$, as a function of temperature, at bias powers fixed at $-83\,\mathrm{dBm}$ (FunKID) and $-68\,\mathrm{dBm}$ (KID)}, chosen to minimize QPs excess~\cite{qp_power, qp_power2, qp_power3, qp_power4}. \rev{In Fig.~\ref{fig:scan_T} we compare the data to the model, calculated with the measured $\alpha$ and $\Delta_0$ previously discussed. However, since the BCS theory might be limited in case of inhomogeneous and multilayer superconductors, the comparison is qualitative and aims to test the applicability of the model.}

In both cases, data deviate from the BCS model. The FunKID shows significant discrepancies at intermediate temperatures, indicating that the presence of the funnels and the trilayer KID structure prevents the BCS model from fully describing the device. For the normal KID, the overall behavior of data is similar to the BCS model, although deviations at high temperatures are still observed. Therefore more complete models than the BCS-based one would be required for a more accurate interpretation~\cite{zhao2017}, but these are beyond the scope of this work.

To calibrate the detectors, the FunKID and the KID were operated at bias powers $P_\mathrm{{cal}}$ of \rev{-65\,dBm and -59\,dBm}, respectively, at which they exhibited good performance in terms of signal to noise ratio~\cite{altial}. Optical pulses \rev{releasing in the substrate} increasing total energy from around 9\,keV up to 90\,keV were fired onto the back of the substrate by means of the central fiber~\cite{altial,KID_germanium, lantern}. Finally, signals were processed offline with a matched filter~\cite{matched_filter} in order to improve the signal to noise ratio. \rev{The same procedure was repeated to calibrate the detectors shining with the fibers aligned with the FunKID and the KID, respectively.}

\begin{figure}[t]
    \centering
    \includegraphics[width=\columnwidth]{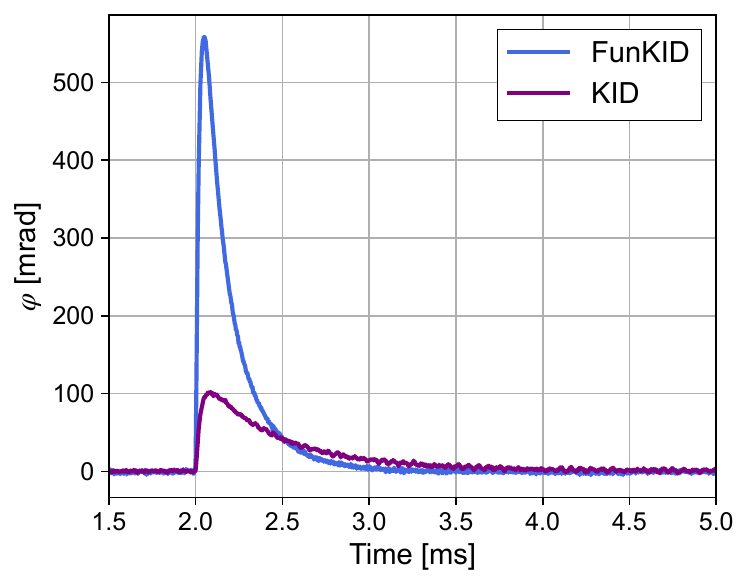}
    \caption{Two pulses generated by the same energy release \rev{ firing with the central fiber}. A higher response is observed for the FunKID resonator compared to the normal KID.}
    \label{fig:pulse}
\end{figure}

Two pulses generated by \rev{the same} energy release of around 90\,keV into the substrate \rev{fired by the central fiber} are superimposed in Fig.~\ref{fig:pulse}. \rev{The FunKID exhibits a pulse response about 5.5 times larger than the standard KID. The decay times $\tau_d$ of the FunKID and KID are about 0.15\,ms and 0.45
\,ms, respectively. The smaller decay time of the FunKID is in line with the smaller internal quality factor~\cite{calder2}. The rise times $\tau_{\mathrm{rise}}$ amount to around 12\,$\mathrm{\mu s}$ and 20\,$\mathrm{\mu s}$, respectively. These are significantly \rev{longer} than the respective ring times (see Tab~\ref{tab:table1}). It is reasonable to believe that in both cases the rise times of the pulses are dominated by the phonon \rev{arrival time} $\tau_{\mathrm{ph}}$~\cite{ eta_phonon}.}

\begin{table*}[t]
\centering
\begin{tabular}{c|c|c|c|c}
\hline
\hline

 & \multicolumn{2}{c|}{\textbf{FunKID}} & \multicolumn{2}{c}{\textbf{KID}} \\ \cline{1-5} 

 Fiber & \rev{F} & C & \rev{K} & C \\ \hline

$P_\mathrm{{cal}}\,\mathrm{[dBm]}$ & -\rev{65}& -65& \rev{-59}& -59\\ 
$r\,\mathrm{[mrad/keV]}$  & \rev{$6.78\pm0.19$}& $5.85\pm 0.14$& \rev{$1.88\pm 0.07$}& $1.11\pm 0.04$\\ 
\rev{$\sigma_0 \,\mathrm{[mrad]}$} &\rev{$0.372 \pm 0.009$ }& $0.386 \pm 0.009$ & \rev{$0.165 \pm 0.005$} & $0.173 \pm 0.005$\\ 
$\sigma_E\,\mathrm{[eV]}$ & \rev{$55.0 \pm 2.0$}& $65.9 \pm 2.3$& \rev{$88.5 \pm 3.9$} &  $156 \pm 6$\\ 
\hline
\hline
\end{tabular}
\caption{Results of the energy calibrations performed on the two KIDs. For each device, we report the calibration power $P_{\mathrm{cal}}$, the phase responsivity $r$, the noise RMS after the matched filter $\sigma_0$ and energy resolution $\sigma_E$. \rev{ All the quantities are measured with the central fiber C and with the closest fiber to each resonator (fiber F for the FunKID, fiber K for the KID).}}
\label{tab:tot_results}
\end{table*}

\rev{The results of the energy calibrations are reported  in Tab.~\ref{tab:tot_results} (the results for the energy calibration with the central fiber are discussed also in the supplementary material). The responsivity obtained with the central fiber is about 5.3 higher at the calibration power for the FunKID, that is in line with the higher pulse amplitude shown in Fig.~\ref{fig:pulse} and previously discussed. Due to the higher noise RMS of the FunKID, the enhanced responsivity only translates into a milder improvement of the energy resolution, corresponding to a factor $\sim2$. Due to position effects, the ratio of the responsivities is smaller if we shine each detector with the fiber aligned with it amounting to $\sim 3.6$.

The efficiency ratio $\eta_\mathrm{F}/\eta_\mathrm{K}$ is obtained by plugging in Eq.~\ref{eq:gain} the measured values of the responsivities, $Q$ and $\alpha$ of the detectors, yielding:
\begin{equation}\notag
    \frac{\eta_\mathrm{F}}{\eta_\mathrm{K}}\simeq \left(\frac{r_{\mathrm{FunKID}}}{r_{\mathrm{KID}}}     \frac{\alpha_{\mathrm{KID}}}{\alpha_{\mathrm{FunKID}}}
    \frac{Q_{\mathrm{KID}}}{Q_{\mathrm{FunKID}}}-1\right) 
    \frac{\Delta_{\mathrm{Al}}}{\Delta_{\mathrm{AlTiAl}}}
\end{equation}
\begin{equation}
           = \left(3.6\times\frac{15.6}{18.9}\times\frac{19}{10}-1\right)\times\frac{1.2}{0.8}\sim 7
\end{equation}

 This value is in line with the ratio of the surface $A_\mathrm{ph}/A_\mathrm{KID}= 9.0$. proving that an higher phonon collection efficiency is achieved. Possible discrepancies with the exact surface coverage ratio may come from QPs losses from the funnels to the resonator, due to the non-perfect trapping efficiency. It is worth to highlight that a possible source of systematic uncertainty on this result arises from the modeling of Eq.~\ref{eq:gain} and from the estimate of $\alpha$.
}

\rev{This work is a proof of concept for the possibility of operating a KID with integrated phonon collectors, which we called FunKID. We observed a successful operation of this new device and an increase in the detector responsivity compared to an identical resonator without collectors. Such increase can be ascribed to the larger phonon collection surface, separated from the active volume of the sensor. However, an enhancement in energy resolution is still required, as the current performance remains worse than previously achieved results~\cite{bullkid_fondo}. To this end, it is crucial to increase the internal quality factor $Q_i$, as well as developing a rigorous model describing QPs dynamics for a better description and possible optimizations of the device.}


\section*{Supplementary materials}
\rev{See the supplementary material for the measurements on the Aluminum devices and details on the calibrations.}

\begin{acknowledgments}
This work was supported by the INFN,  Sapienza University of Rome and co-funded by the European Union (ERC, DANAE, 101087663). Views and opinions expressed are however those of the author(s) only and do not necessarily reflect those of the European Union or the European Research Council. Neither the European Union nor the granting authority can be held responsible for them. We thank G. Catelani for useful discussions. We thank A. Girardi and M. Iannone of the INFN Sezione di Roma for technical support. We acknowledge the support of the PTA
platform for the fabrication of the device. 
\end{acknowledgments}
\section*{Author declarations}
\subsection*{Conflict of Interest}
The authors have no conflicts to disclose.
\section*{Data Availability Statement}
The data that support the findings of this study are available from the corresponding author upon reasonable
request.
\bibliography{bibliography}

\end{document}